\documentclass [preprint,aps,showpacs]{revtex4}
\usepackage[final]{graphics}
\usepackage{amssymb}
\usepackage{amsfonts}
\usepackage{epsfig}
\usepackage{graphicx}

\begin{document}

\begin{center}
{\Large Phase diagram and critical behavior of the
\vspace{.5em}

pair annihilation model}
\end{center}
\vspace{1cm}

\centerline{\large Adriana Gomes Dickman$^{1,*}$ and Ronald Dickman$^{2,\dagger}$}
\begin{center} $^1$Departamento de F\'\i sica e Qu\'\i mica, Pontif\'\i cia
Universidade Cat\'olica de Minas \\ Gerais, Av. Dom Jos\'e Gaspar, 500,
Cora\c c\~ao Eucar\'\i stico, 30535-901, Belo Horizonte, Minas Gerais, Brazil     \\
\vspace{1em}

$^2$Departamento de F\'\i sica, ICEx, and Universidade Federal de Minas Gerais \\
 Caixa Postal 702, 30123-970, Belo Horizonte, Minas Gerais, Brazil,\\
 National Institute of Science and Technology for Complex Systems,\\
Caixa Postal 702, 30161-970 Belo Horizonte, Minas Gerais,
Brazil
\end{center}

\date{\today}

\noindent We study the critical behavior of the pair annihilation
model (PAM) with diffusion in one, two and three dimensions,
using the pair approximation (PA) and Monte Carlo simulation.
Of principal interest is the dependence of the critical creation rate,
$\lambda_c$, on the diffusion probability $D$, in particular, whether survival
is possible at arbitrarily small creation rates, for sufficiently rapid diffusion.
Whilst the PA predicts that in any spatial dimension $d \geq 1$, $\lambda_c \to 0$ at some diffusion
probability $D^* < 1$,
Katori and Konno [Physica A {\bf 186}, 578 (1992)] showed rigorously
that for $d \leq 2$, one has $\lambda_c > 0$ for any $D<1$.
Our simulation results are consistent with this theorem.
In two dimensions, the extinction
region becomes narrow as $D$ approaches unity,
following $\lambda_c \propto \exp[- \rm{const.}/(1-D)^\gamma]$,
with $\gamma = 1.41(2)$.  In three dimensions we find $D^* = 0.333(3)$.
The simulation results confirm that the PAM belongs to the directed
percolation universality class.

\vspace{3.0truecm}


\vspace{1.0truecm}

\noindent {\small electronic address: $^*$adickman@pucminas.br} \\
\noindent {\small electronic address: $^\dagger$dickman@fisica.ufmg.br} \\

\newpage

\section{Introduction}

In recent decades, general theories of phase transitions and critical phenomena
have been developed, unifying our understanding of equilibrium phase transitions
in liquid-vapor, magnetic, liquid crystals and other systems. By contrast, the
study of nonequilibrium critical phenomena is still in development. Since the
transition rates in such systems do not satisfy detailed balance, the
steady-state probability distribution in these systems is not known a priori,
and the analysis must be based upon the dynamics.
Starting from the basic contact process \cite{harris}, many particle
systems have been studied in efforts to characterize scaling properties at nonequilibrium
phase transitions \cite{marro,henkel,odor}.  These models, which involve creation and annihilation of
particles on a lattice,
typically exhibit a phase transition to an {\it absorbing state}
(one allowing no escape), and so violate the detailed balance principle.
An issue that has attracted some interest is the combined effect of multiparticle rules
and diffusion (hopping), which tends to spread particles uniformly over the system.

In this work we revisit the pair annihilation
model (PAM) \cite{pam1,pam2}.
In this model particles diffuse on a lattice at rate $D$, nearest-neighbor pairs of
particles are annihilated at rate $(1-D)/(1+\lambda)$, and particles attempt to create new particles at rate
$(1-D)\lambda/(1+\lambda)$.
Double occupancy of sites is forbidden.
The model exhibits active and absorbing phases,
separated by a continuous phase transition at $\lambda_c(D)$.
Using cluster approximations and Monte Carlo simulation, we determine the phase boundary
in one, two, and three dimensions.

The pair approximation predicts that for a diffusion rate
greater than a certain value, $D^*$, the critical parameter $\lambda_c=0$.
(That is, for $D>D^*$, an arbitrarily small creation rate is sufficient to
maintain a nonzero particle density.)
This prediction is known to be wrong in dimensions $d \leq 2$: Katori and Konno \cite{katori} proved
that $\lambda_c > 0$ for any diffusion probability $D<1$, in one and two dimensions.
Their theorem is based on a relation between the PAM and the branching annihilating
random walk of Bramson and Gray \cite{gray}.
Existence of a $D^* < 1$ is not ruled out in $d \geq 3$ dimensions.  The difference
is connected with the nonrecurrence of random walks in $d \geq 3$ \cite{katori}.
How $\lambda_c$ tends to zero as $D \to 1$ is, however, unknown.
Moreover the question of whether, in three or more dimensions, $D^*$ is in fact less that unity,
has not, to our knowledge, been studied.
The principal motivation for the present work is
to determine $\lambda_c(D)$ via numerical simulation.
We also verify that the model belongs to the directed percolation (DP) universality class,
as expected on the basis of symmetry considerations \cite{jans81,gras82}.
Our simulation results, while consistent
with the Katori-Konno theorem, show that in the two-dimensional case, $\lambda_c$ becomes extremely
small as $D$ approaches unity, possibly suggesting the (incorrect)
impression that the critical value is actually zero
at some finite diffusion rate.

The remainder of this paper is organized as follows.  In the following section (II) we define the
model and discuss its limiting behaviors in the $\lambda$-$D$ plane.  Then in Sec. III we present,
for completeness, the one- and two-site cluster approximations.  Simulation results are discussed
in Sec. IV, followed by a brief discussion in Sec. V.

\section{The model}

The PAM is defined on a lattice,
in which sites can be either occupied by a particle or vacant \cite{marro,pam1,pam2};
we denote these states by $\sigma_x = 1$ (site $x$ occupied) and $\sigma_x = 0$ (site $x$ vacant).
There are three kinds of transition: nearest-neighbor (NN) hopping (``diffusion"), creation,
and pairwise annihilation, with associated rates
$D$, $(1-D)\lambda/(1+\lambda)$, and $(1-D)/(1+\lambda)$, respectively.
(Since the rates are parameterized so as to sum to unity, we are free to refer to $D$ as the
diffusion {\it probability}.)
At each step of the evolution, the next attempted transition is taken as diffusion, creation, or
annihilation, with probabilities
$D$, $(1-D)\lambda/(1+\lambda)$, and $(1-D)/(1+\lambda)$, respectively.

In a hopping transition,
a site $x$ is chosen at random, along with a nearest-neighbor site $y$ of $x$.
Then if $\sigma_x \neq \sigma_y$, the states are exchanged.  In a creation event, a site $x$ is
chosen. If $\sigma_x =1$, a nearest-neighbor $y$ is chosen, and if $\sigma_y = 0$ this variable
is set to one.  Finally, in an annihilation event, a pair of nearest-neighbor sites $x$ and $y$
are chosen at random, and if $\sigma_x = \sigma_y = 1$, both variables are set to zero.  Each
transition corresponds to a time increment $\Delta t = 1/N_{site}$, where $N_{site}$ is the
number of lattice sites.

To improve efficiency, in simulations the
site $x$ is chosen from a list of occupied sites;
then the time increment is $\Delta t = 1/N_p$, with $N_p$ the number of {\it particles} in the
system, immediately prior to the transition.  In this implementation, the rate of annihilation
of a given NN particle pair is
\begin{equation}
R_{an} = \frac{1}{\Delta t} \frac{1-D}{1+\lambda} \frac{2}{N_p} \frac{1}{2d}
= \frac{1}{d} \frac{1-D}{1+\lambda}
\label{effannrate}
\end{equation}
where the factor $2/N_p$ arises because either particle in the pair can be selected from the
list of $N_p$ particles.

The particle-free configuration is absorbing.  By analogy with the contact process \cite{marro,harris},
we expect that in the infinite-size limit the system undergoes a phase transition between
an active state (with nonzero stationary particle density) and an absorbing one, as one crosses
the critical line $\lambda_c(D)$ in the $\lambda$-$D$ plane.  As creation depends upon a
single particle, the order parameter is the
stationary density of particles, $\rho$.

When a new particle is created, it
always forms a pair with the ``parent" particle, making these two particles susceptible to annihilation.
In the active stationary state, increasing $D$ at fixed $\lambda$ tends to reduce the fraction of
nearest-neighbor particle pairs toward its random mixing value, $\rho^2$.  Thus we should expect
$\lambda_c$ to be a decreasing (or at least, nonincreasing) function of $D$.
In the simplest mean-field theory analysis, the annihilation rate is proportional to $\rho^2$, so that
for small $\rho$, one has $\dot{\rho} \propto \lambda \rho -  \mbox{const.} \times \rho^2 $, which admits a
stationary solution $\rho \propto \lambda$ for {\it any} nonzero creation rate.  In the limit $D \to 1$
we expect mean-field theory to hold, so that $\lambda_c \to 0$ in this limit.  This raises the question
of whether $\lambda_c$ vanishes at some diffusion
probability $D^*$ strictly less than unity.  While the two-site approximation predicts $D^* < 1$ in any
dimension, the
results of Katori and Konno \cite{katori} imply that $D^* = 1$ in dimensions $d \leq 2$.

The phase diagram of the PAM is expected to have the form shown in Fig. 1.  For $D < D^*$
the behavior along the critical line $\lambda_c (D)$ should be that of DP, given that such
behavior is generic for absorbing-state phase transitions without special symmetries or conserved
quantities \cite{jans81,gras82}.  If $D^* < 1$, then we expect mean-field
like critical behavior as $\lambda \to \lambda_c = 0$ at fixed $D > D^*$.  Within the absorbing phase, for
$0 < \lambda < \lambda_c (D)$, an isolated particle can produce an offspring, leading to annihilation
of both the original and the new particle.  On the line $\lambda = 0$, this channel to
annihilation is not available, and isolated particles cannot disappear.  Thus the dynamics at long
times, for $D>0$, will be that of the diffusive annihilation process $A + A \to 0$, for which the particle
density $\rho(t)$ decays $\sim 1/\sqrt{t}$ in $d=1$,
$\sim (\ln t)/t$ in two dimensions,
and $\sim 1/t$ in $d \geq 3$
\cite{torney,benav}.  Finally, at the point $\lambda = D = 0$, starting from
all sites occupied, pairs are annihilated successively until only isolated particles remain.  This
is equivalent to the random sequential adsorption (RSA) of dimers.  (In the present case, of course,
dimers are removed not adsorbed, so the final particle density is $1 - 2\theta_\infty$, where
$\theta_\infty$ is the final coverage in RSA.)
On the line, the final density
of isolated particles is $e^{-2} = 0.135335...$ \cite{flory}, while in two dimensions one has
$\rho_\infty \simeq 0.093108(8)$ \cite{dimerRSA}.  One may anticipate interesting crossover behaviors in the
vicinity of one or another limit.  In the present work, however, we focus on determining
the function $\lambda_c (D)$ using Monte Carlo simulation.

\begin{figure}[!h]
\epsfysize=9cm
\epsfxsize=10cm
\centerline{\epsfbox{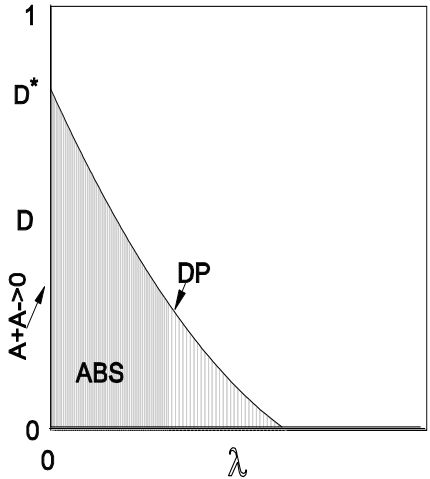}}
\caption{\sf Schematic phase diagram of the PAM in the $\lambda$-$D$ plane.
The results of \cite{katori} imply that $D^*=1$ in one and two dimensions.}
\label{fig:phase}
\end{figure}

\section{Cluster approximations}

In this section we study the PAM through mean-field -- site and pair approximations \cite{ben}.
In general, mean-field results provide a good qualitative description of the phase
diagram and give an order-of-magnitude estimate of the critical point.
$n$-site approximations are a natural way to improve the mean-field approach. The method
consists of treating the transitions inside clusters of $n$ sites exactly, while
transitions involving sites outside the cluster are treated in an approximate manner.

\vspace{0.5cm}
\subsection{One-site approximation}
\vspace{0.5cm}

Let $\rho = \mbox{Prob}(\sigma_x = 1)$ denote the density of particles.  The density is governed by,
\begin{eqnarray}
\frac{d\rho}{dt} &=& \frac{1}{2d}(1-D)\frac{\lambda}{1+\lambda}
\sum_{\hat{e}} P(\sigma_x=0, \sigma_{x+\hat{e}}=1)
\nonumber \\
&-& \frac{1}{d} \frac{1-D}{1+\lambda} \sum_{\hat{e}} P(\sigma_x=1, \sigma_{x+\hat{e}}=1)
\nonumber \\
&-&D\sum_{\hat{e}} P(\sigma_{x}=1, \sigma_{x+\hat{e}}=0)
\nonumber \\
&+&D\sum_{\hat{e}} P(\sigma_{x}=0, \sigma_{x+\hat{e}}=1),
\label{eq:ee}
\end{eqnarray}
where the sums are over the $2d$ nearest-neighbors of site $x$, and
$P(\sigma_{x},\sigma_{x+\hat{e}})$ is a two-site joint probability.
Equation~(\ref{eq:ee}) couples the one-site probability $\rho$ to the
two-site probabilities, which in turn
depend on the three-site probabilities, and so forth, leading to an infinite
hierarchy of equations for the $n$-site probabilities. The site-approximation
consists in truncating this hierarchy at $n = 1$, so that the two-site
probabilities are replaced by a product of two one-site
probabilities. Assuming spatial homogeneity and isotropy we obtain the following
equation for $\rho$,
\begin{eqnarray}
{d\rho \over dt} = \frac{1-D}{1+\lambda} \left[ \lambda\rho -
                           (2+\lambda)\rho^2 \right]
\label{eq:sa}
\end{eqnarray}
The stationary solutions are $\overline\rho=0$ (unstable for $\lambda>0$) and
$\overline\rho=\lambda/(2 + \lambda)$. Thus, in this approximation the critical parameter $\lambda_c$ is
zero in any dimension. Notice that in this approximation the diffusion rate has no
influence on the stationary solution.

\vspace{0.5cm}
\subsection{Pair approximation}
\vspace{0.5cm}


To derive the pair approximation equations, we consider the changes in the configuration of a
NN pair of sites (the {\it central} pair), given the states of the surrounding sites.
Using the symbols $\circ$ and $\bullet$ to represent, respectively, vacant and occupied sites,
the states of a pair are $\circ\circ$, $\bullet\bullet$, $\bullet\circ$, and $\circ\bullet$;
the latter two have the same probability and may be treated as a single class using appropriate
symmetry factors.  It is convenient to use $(\bullet\bullet)$ to denote the fraction of
$\bullet\bullet$ pairs, and so on.  Then we have for the site fractions $(\bullet) = \rho$ and
$(\circ) = 1 -\rho$:

\begin{eqnarray}
(\bullet) &=& (\bullet\bullet) + (\bullet\circ),\\
(\circ)   &=& (\circ\circ)     + (\bullet\circ).
\label{eq:concentration}
\end{eqnarray}
\noindent The pair fractions satisfy $(\circ\circ) + 2 (\circ\bullet) + (\bullet\bullet) = 1$.
The pair approximation consists in writing the joint probability of a set of three neighboring
sites in the form $(abc) = (ab)(bc)/(b)$.

There are five possible transitions between the pair states.  Consider for example
the transition $\circ\circ \to \circ\bullet$.  This can occur via creation or via
hopping, if and only if the rightmost site of the central pair has an occupied
NN.  Since its NN {\it within} the central pair is vacant,
at least one of its $2d-1$ NNs {\it outside} the central pair
must be occupied.  The rate of transitions via creation is
\begin{equation}
R_{1,c} = (1-D) \tilde{\lambda} \frac{2d-1}{2d} \frac{(\circ\circ)(\circ\bullet)}{(\circ)}
\label{r1c}
\end{equation}
where we introduced $\tilde{\lambda} = \lambda/(1+\lambda)$.  Adding the contribution due to diffusion,
we obtain the total rate for this transition,
\begin{equation}
R_{1} = \frac{2d-1}{2d} \frac{(\circ\circ)(\circ\bullet)}{(\circ)} [D + (1-D) \tilde{\lambda}]
\label{r1c}
\end{equation}
Note that the contribution to the loss term for $(\circ\circ)$ associated with this process is $2R_1$,
due to the mirror transition $\circ\circ \to \bullet\circ$.

The rates for the other transitions are:

$\circ\bullet \to \circ\circ$:
\begin{equation}
R_2 = \frac{2d-1}{2d} \frac{(\circ\bullet)}{(\bullet)} [D (\circ\bullet)
+ 2 (1-D)(1- \tilde{\lambda}) (\bullet\bullet)]
\end{equation}

$\circ\bullet \to \bullet\bullet$:
\begin{equation}
R_3 = \frac{2d-1}{2d} \frac{(\circ\bullet)^2}{(\circ)} [D
+ (1-D)\tilde{\lambda}] + \frac{1}{2d} (1-D) \tilde{\lambda} (\circ\bullet)
\end{equation}

$\bullet\bullet \to \circ\circ$:
\begin{equation}
R_4 = \frac{1}{d} (1-D) (1- \tilde{\lambda}) (\bullet\bullet)
\end{equation}

$\bullet\bullet \to \circ\bullet$:
\begin{equation}
R_5 = \frac{2d-1}{2d} \frac{(\bullet\bullet)}{(\bullet)} \left[2 (1- D) (1- \tilde{\lambda}) (\bullet\bullet)
+ D (\circ\bullet) \right]
\end{equation}

The equations of motion for the pair probabilities are then

\begin{equation}
\frac{d}{dt} (\circ\circ) = 2R_2 + R_4 - 2R_1
\end{equation}

\begin{equation}
\frac{d}{dt} (\circ\bullet) = R_1 + R_5 -R_2 -R_3
\end{equation}
and
\begin{equation}
\frac{d}{dt} (\bullet\bullet) = 2R_3 - R_4 - 2 R_5
\end{equation}
\vspace{1em}

\noindent The active stationary solution of the above equations is
\begin{equation}
\overline{\rho} = \frac{\lambda[(4d-3+D)\lambda - 2(1-2dD)]}
{(4d-3+D)\lambda^2 + 2[2d(D+2) -3]\lambda + 4(2d-1)D},
\end{equation}
and
\begin{equation}
\overline{(\bullet\bullet)} = \frac{\lambda}{\lambda +2} \, \overline{\rho} \,.
\end{equation}
\vspace{1em}

\noindent For $ \lambda < 2(1-2dD)/(4d-3+D)$, only the trivial solution ($\overline{\rho} =0$) exists.
If $D \geq D^* = 1/2d$, however, the active solution exists for any $\lambda > 0$.
The phase transition occurs at
\begin{equation}
\lambda_c = \left\{ \begin{array} {cc} \frac{2(1-2dD)}{4d-3+D}, \;\;\;\;\;\;
D < D^* = \frac{1}{2d} \\
\\
0, \;\;\;\;\;\; D > D^*

\end{array} \right.
\end{equation}

\noindent Thus the pair approximation predicts a nonzero critical creation rate
only for diffusion probabilities $D < D^* = 1/(2d)$; for larger values of $D$, there is
a nonzero particle density for any $\lambda > 0$, as in the one-site approximation.
For $D=0$, we have $\lambda_c = 2$, 2/5, and 2/9 in one, two and three
dimensions, respectively; the corresponding values from simulation are $\lambda_c = 5.368(1)$,
1.0156(1), and 0.475(1).
[We note that the pair approximation results derived above differ slightly from those
given in \cite{marro} since in the latter case the annihilation rate for a NN
particle pair is taken as $(1-D)/(1+\lambda)$, i.e., $d$ times the rate given in
Eq. (\ref{effannrate}).]

Katori and Konno \cite{katori} proved that the prediction of $D^* < 1$, furnished
by the pair approximation, is wrong for $d\leq 2$. That is,
in one and two dimensions, $\lambda_c > 0$ for any $D<1$.
In the following section we investigate
how $\lambda_c$ tends to zero as $D \to 1$ in one and two dimensions, and determine
$D^*$ in the three-dimensional case.

\section{Simulations}

We use Monte Carlo simulations to obtain accurate values of
the critical creation rate $\lambda_c(D)$ and the critical exponents of the PAM in one, two,
and three dimensions.

\subsection{One dimension}

\subsubsection{Spreading behavior}

A well established method for determining the critical point and certain critical exponents
is through the study of propagation of activity, starting from a localized seed, as
proposed long ago by Grassberger and de la Torre \cite{grassberger}.
One studies the activity in a large set
of trials, all starting from a configuration very close to the absorbing
state. Here the initial configuration is that of a single pair of particles at the two central sites,
in an otherwise empty lattice.  Each trial ends when it reaches
the absorbing state, or at a
maximum time, $t_{max}$, chosen such that the activity never reaches the edges
of the system (in any trial) for $t \leq t_{max}$.

For $\lambda>\lambda_c$
there is a nonzero probability that the process survives as $t\rightarrow
\infty$; for $\lambda \leq \lambda_c$ the process dies with probability 1. Of
primary interest are $P(t)$, the probability of surviving until time $t$ or greater, $n(t)$,
the mean number of particles (averaged over all trials), and $R^2(t)$, the
mean-square distance of particles from the origin. At the critical point
these quantities follow asymptotic power laws,
\begin{eqnarray}
P(t)&\propto& t^{-\delta} \\
n(t)&\propto& t^\eta \\
R^2(t)&\propto& t^{z_{sp}} .
\end{eqnarray}
The exponents $\delta$, $\eta$, and $z_{sp}$ satisfy the hyperscaling relation $4\delta+2\eta=dz_{sp}$, in
$d\leq 4$ dimensions \cite{grassberger}.  (We note that $z_{sp}$ is related to the usual dynamic
exponent $z$ via $z_{sp} = 2/z$.)

We study activity spreading in one dimension using samples
of from $10^6$ or $2 \times 10^6$ trials for each $\lambda$ value of interest.
The maximum time $t_{max}$ = 15$\,$000 for $D \leq 0.7$, 30$\,$000 for $D=0.8$ and 0.9, and
$60\,000$ for $D=0.95$.  (As $D$ increases, the asymptotic power-law behavior
occurs at ever later times.)  To ensure that activity never reaches the borders, we use
a lattice size of $L=50\,000$ for $t_{max}$ = 15$\,$000, and $L=80\,000$ for the
longer studies.  A study performed at a given value of $\lambda$ is used to generate
results for nearby values using sample reweighting \cite{reweighting}.

\begin{figure}
\epsfysize=12cm
\epsfxsize=14cm
\centerline{\epsfbox{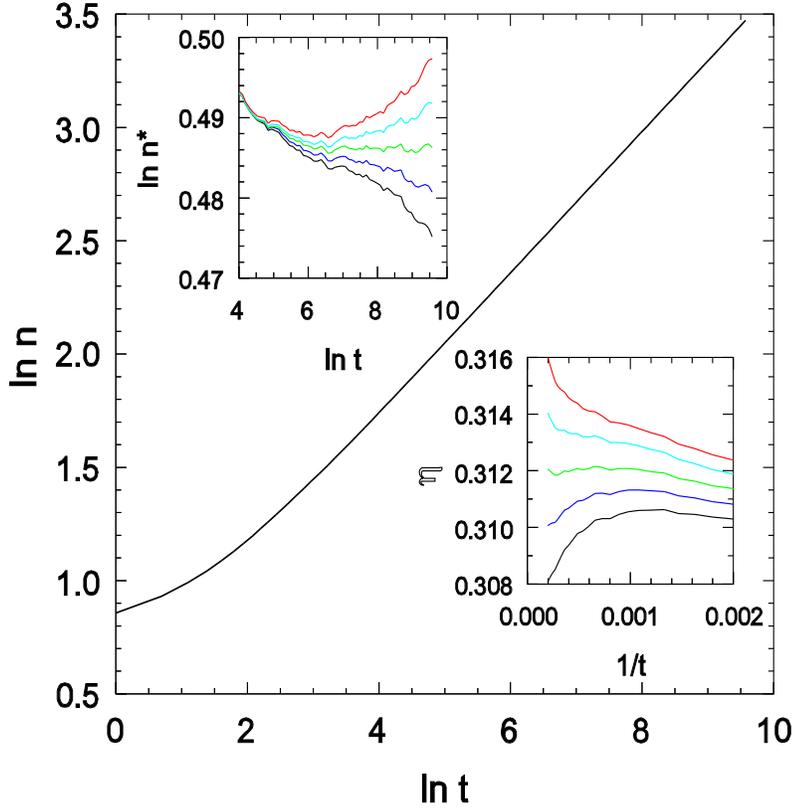}}
\caption{\sf Main graph: $n(t)$ on log scales for the one-dimensional PAM
with $D=0.3$ and $\lambda=3.4687$.  Upper inset: $n^* = n/t^\eta$ on log scales, for
(lower to upper) $\lambda=$ 3.4681, 3.4684, 3.4687, 3.4690, and 3.4693. Lower inset:
local slopes $\eta(t)$ for the same set of $\lambda$ values.}
\label{fig:spread}
\end{figure}

To locate the critical point, we use the criterion of power-law behavior of $n(t)$;
Fig.~\ref{fig:spread} illustrates the analysis for $D=0.3$.  The main graph is a log-log plot of
$n(t)$ showing an apparent power law for $\lambda=3.4687$.  The curves for nearby values
(specifically, $\lambda$ = 3.4681, 3.4684, 3.4690, and 3.4693, obtained via reweighting),
cannot be distinguished on the scale of this graph, but if we plot $n^* \equiv n/t^\eta$, the curves for
different $\lambda$ values fan out (upper inset), with upward curvature
indicating a supercritical value of $\lambda$ and vice-versa.

The exponent $\eta$
is estimated via analysis of the
local slope, $\eta(t)$,
defined as the inclination of a least-square linear fit to the data (on logarithmic scales),
on the interval $[t/a, \,at]$.  (The choice of the factor $a$ represents a compromise between
high resolution, for smaller $a$, and insensitivity to fluctuations, for larger values;
here we use $a = 2.59$.)  Plotting $\eta(t)$ versus $1/t$ (lower inset of
Fig.~\ref{fig:spread}) allows one to estimate $\lambda_c$ (the curves for $\lambda>\lambda_c$
veer upward, and vice-versa), and to estimate the critical exponent $\eta$ by
extrapolating $\eta(t)$ to $1/t\rightarrow 0$.
The main source of uncertainty in the exponent estimates is the uncertainty in $\lambda_c$
itself.
An analogous procedure is used to estimate
exponents $\delta$ and $z_{sp}$.
In Table~\ref{tb:pamsim1d} we list the critical parameters and spreading exponents
found via spreading simulations combined with local-slopes analysis.

\begin{table}[!h]
\begin{center}
\begin{tabular}{|c|c|c|c|c|} \hline
$D$ & $\lambda_c$ & $\delta$ & $\eta$     & $z_{sp}$       \\ \hline\hline
0.0 & 5.3720(5)   & 0.159(1) & 0.315(1)   & 1.266(2) \\
0.1 & 4.6709(2)   & 0.161(1) & 0.314(2)   & 1.264(1) \\
0.2 & 4.0417(2)   & 0.160(1) & 0.314(1)   & 1.266(1) \\
0.3 & 3.4687(2)   & 0.162(1) & 0.312(1)   & 1.268(2) \\
0.4 & 2.9411(2)   & 0.160(2) & 0.314(2)   & 1.264(2) \\
0.5 & 2.4473(1)   & 0.159(1) & 0.315(2)   & 1.267(3) \\
0.6 & 1.9778(2)   & 0.159(1) & 0.315(2)   & 1.266(1) \\
0.7 & 1.5231(2)   & 0.158(2) & 0.315(1)   & 1.265(2) \\
0.8 & 1.0684(2)   & 0.159(2) & 0.315(2)   & 1.265(2) \\
0.9 & 0.5891(1)   & 0.161(1) & 0.315(1)   & 1.267(3) \\
0.95& 0.3214(1)   & 0.159(2) & 0.318(3)   & 1.266(4) \\ \hline
\end{tabular}
\end{center}
\caption{\sf Results of spreading simulations for the PAM in one dimension.}
\label{tb:pamsim1d}
\end{table}

For $D=0$, the critical parameter for the PAM,
$\lambda_c (0)$=5.368(1), is considerably larger than that of the contact process
($\lambda_c$=3.29785(2)), as expected since here each annihilation event removes
two particles.  (The fact that $\lambda_c$ is {\it less than twice} the corresponding
value in the CP may be attributed to the tendency for particles to cluster: removing two
particles may eliminate additional pairs, thus reducing the effective rate of
annihilation.)

\begin{figure}[h]
\epsfysize=9cm \epsfxsize=12cm \centerline{ \epsfbox{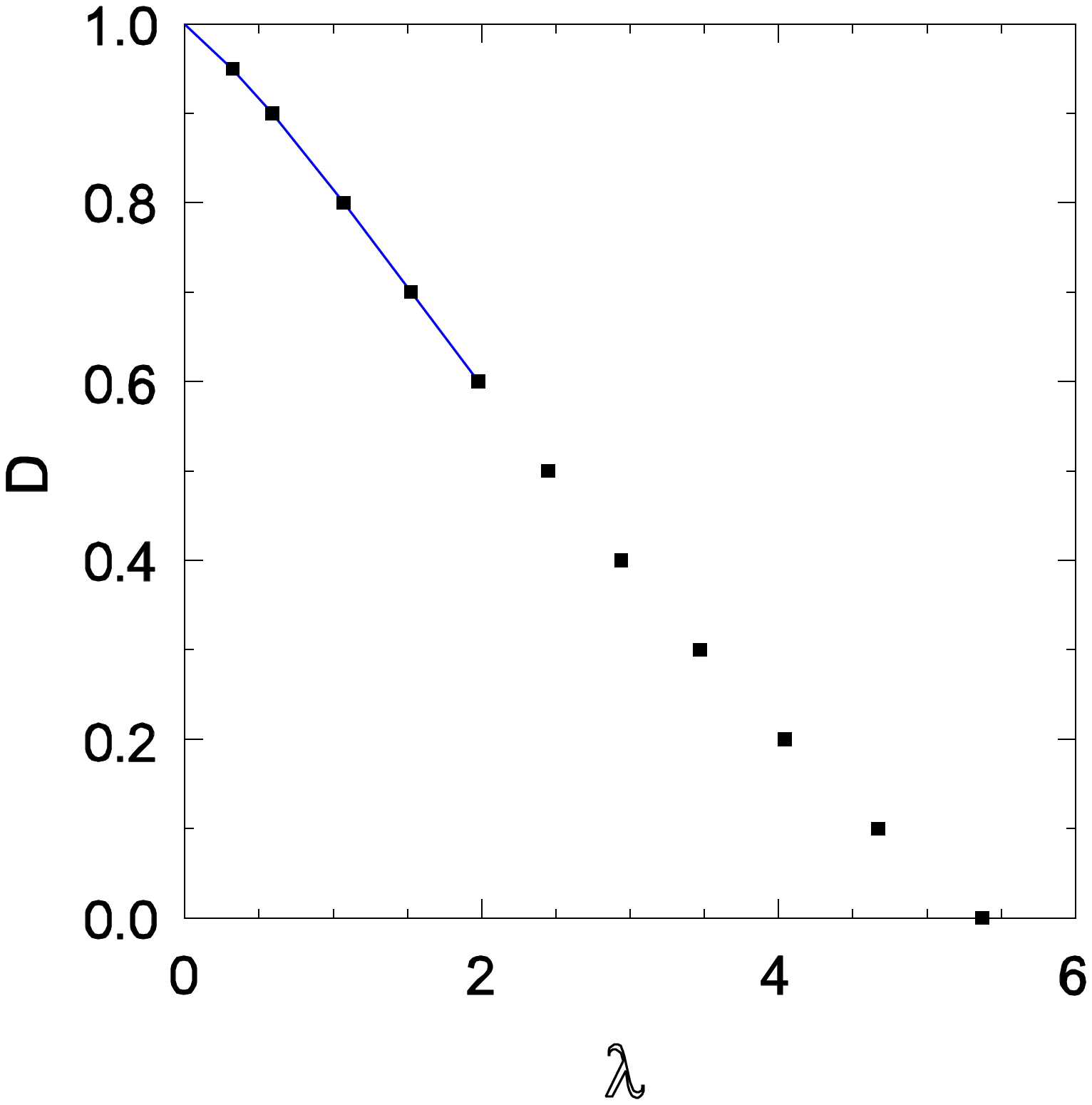}}
\vspace{-3em}

\caption{\sf Points along the critical line in the $\lambda$-$D$ plane in one dimension,
as determined via simulation.  Error bars are smaller than symbols.
The solid line is a quartic fit to the six points with largest $D$.
}
\label{fig:sim1d}
\end{figure}

For all diffusion probabilities studied, our estimates for the critical exponents are
in good accord with the DP values $\delta=0.15947(3)$,
$\eta=0.31368(4)$, and $z=1.26523(3)$ \cite{marro}.
A plot of the phase boundary in the $\lambda$-$D$ plane (see Fig.~\ref{fig:sim1d}) suggests that
$\lambda_c \to 0$ as $D \to 1$, so that $D^* = 1$ in agreement with the
Katori-Konno theorem.  Extrapolation of $D$ versus $\lambda_c$, using a fourth-order polynomial
fit to the data for $D \geq 0.6$, yields $D = 1.0005$ for $\lambda_c=0$, confirming to high
precision that $\lambda_c > 0$ for $D < 1$.

$\;$
\vspace{2em}

\subsection{Two dimensions}

\subsubsection{Steady-state behavior}

We encounter rather large uncertainties in studies of spreading behavior
of the two-dimensional PAM, and so turn to the steady-state approach to investigate
this system (on the square lattice).
In these studies we initialize the system with all sites occupied, and
allow it to evolve until it attains a quasistationary (QS) regime,
in which bulk properties such as the particle density $\rho$, averaged
over surviving realizations, are time-independent.
According to the finite-size scaling hypothesis \cite{fisher,barber},
the QS properties depend on system size $L$ through the ratio $L / \xi$, or equivalently
through the scaling variable $\Delta L^{1/\nu_\bot}$, where $\Delta\equiv \lambda-\lambda_c $.
Expressing the order parameter as a function of $\Delta$ and $L$, we have
\begin{eqnarray}
\rho (\Delta,L)\propto
L^{-\beta/\nu_\bot}
f(\Delta L^{1/\nu_\bot}).
\label{eq:densdeltaL}
\end{eqnarray}
with $f(x)\propto x^\beta$ as $x\rightarrow\infty$.
At the critical point, $\Delta=0$,
\begin{eqnarray}
\rho (0,L)\propto L^{-\beta/\nu_\bot}.
\label{eq:densdeltaLcp}
\end{eqnarray}
Thus an asymptotic power-law dependence of $\rho$ on $L$ is
a useful criterion for criticality.

We study the QS density as a function of system size
to locate the critical point, using sizes $L=25$, 50, 100,...,800.
The relaxation time varies from
$\tau=800$ for the smallest size, to $\tau=200\,000$ for the largest; the
number of realizations varies from 500 to 10\,000.
Using the power-law criterion, we obtain the estimates for $\lambda_c$ listed in
Table~\ref{tb:pamss2d}.
It is worth mentioning that the values for $\lambda_c$,
for $D=0$ and $D=0.1$ are in
good agreement with those obtained in the preliminary spreading behavior studies.

\begin{table}[h]
\begin{center}
\begin{tabular}{|c|c|c|c|c|} \hline
$D$ & $\lambda_c$  \\ \hline\hline
0.0 & 1.0156(1)    \\
0.1 & 0.7877(1)    \\
0.2 & 0.5890(5)    \\
0.3 & 0.4166(1)    \\
0.4 & 0.2685(5)    \\
0.5 & 0.1462(2)    \\
0.6 & 0.056(1)     \\ \hline
\end{tabular}
\end{center}
\caption{\sf Critical parameters obtained through steady-state simulations
in two dimensions.}
\label{tb:pamss2d}
\end{table}

\begin{figure}[!h]
\epsfysize=10cm
\epsfxsize=13cm
\centerline{
\epsfbox{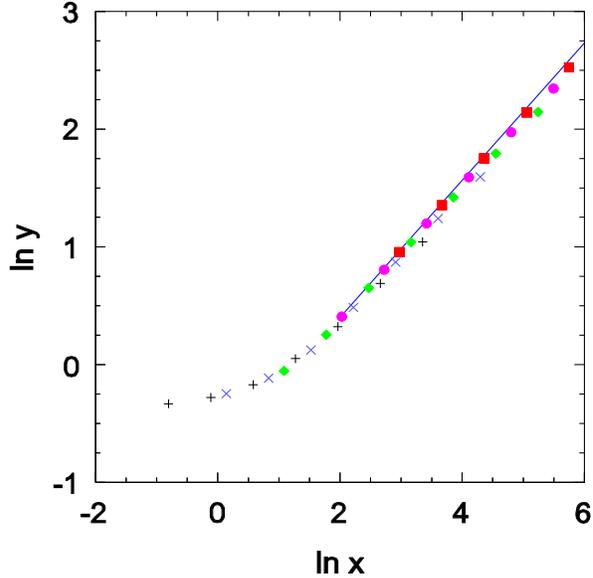}}
\vspace{-2em}

\caption {\sf (Color online)
Scaling plot of the stationary density in the two-dimensional PAM with $D=0$.
System sizes $L$=16 ($+$); 32 ($\times$); 64 (diamonds); 128 ($\bullet$) and 256 (squares).
The slope of the straight line is 0.583.
}
\label{fig:datacol}
\end{figure}

In Fig.~\ref{fig:datacol} we verify the scaling collapse of the order parameter,
plotting $x \equiv L^{\beta/\nu_\perp} \rho$ versus $y \equiv \Delta L^{1/\nu_\perp}$, for system sizes
$L$=16, 32, 64, 128, and 256.  A good collapse is obtained
using the DP values $\nu_\perp$ = 0.733 and $\beta/\nu_\perp = 0.795$
\cite{marro}.
The data are consistent with the scaling law $\rho \propto \Delta^\beta$,
using the DP value $\beta = 0.583(4)$ \cite{marro}.

\subsubsection{Quasistationary simulations}

As $D$ approaches 0.7 the critical value $\lambda_c$ becomes very small.  We require an efficient
simulation method to obtain precise estimates for the critical value for larger diffusion rates,
in particular, to determine how $\lambda_c$ tends to zero as $D$ increases.  For this purpose
we use quasistationary (QS) simulations, which sample directly the QS probability distribution,
that is, the long-time distribution conditioned on survival.  The details of the method are
explained in Ref. \cite{qssim}.  To obtain these results we use lattice sizes $L=100$,
200, 400 and 800 for
$D=0.7$, and include studies of larger systems for higher diffusion rates
(up to $L=6400$, for $D \geq 0.78$).  The critical point is determined
via the criteria of power-law scaling of the density and mean lifetime with system size, and
convergence of the moment ratio $m = \langle \rho^2 \rangle / \rho^2$ to a finite limiting value
as $L \to \infty$, as discussed in \cite{moments}.  (The lifetime $\tau$ is expected
to follow $\tau \sim L^z$.)  Using this method we obtain the values listed in
Table~\ref{tb:qss2d}.  We note that our results for $\beta/\nu_\perp$, $z$, and the limiting moment
ratio $m_c$ are consistent with the known DP values of 0.795(10), 1.7674(6), and 1.3257(5),
respectively \cite{marro,reweighting,moments}.

\begin{table}[h]
\begin{center}
\begin{tabular}{|c|c|c|c|c|} \hline
$D$  & $\lambda_c$  \\ \hline\hline
0.60 & 0.05632(3)   \\
0.70 & 0.00940(5)   \\
0.73 & 0.003957(3)  \\
0.78 & 0.0004815(7) \\
0.80 & 0.00015(2)   \\ \hline
\end{tabular}
\end{center}
\caption{\sf Critical parameters obtained through quasistationary simulations
in two dimensions.}
\label{tb:qss2d}
\end{table}

\noindent For $D=0.8$, $\lambda_c$ is of order $10^{-4}$, and a precise determination becomes very
difficult due to the small number of particles present in the system.  Reliable
determination of $\lambda_c$ for larger diffusion rates would therefore require studies
of even larger systems, which was deemed impractical.

We find that $\lambda_c(D)$ can be fit quite well using an expression of the form

\begin{equation}
\lambda_c = A \exp \left[ - \frac{C}{(1-D)^\gamma}\right]\,.
\label{fit2d}
\end{equation}

\begin{figure}[!h]
\epsfysize=10cm
\epsfxsize=12cm
\centerline{
\epsfbox{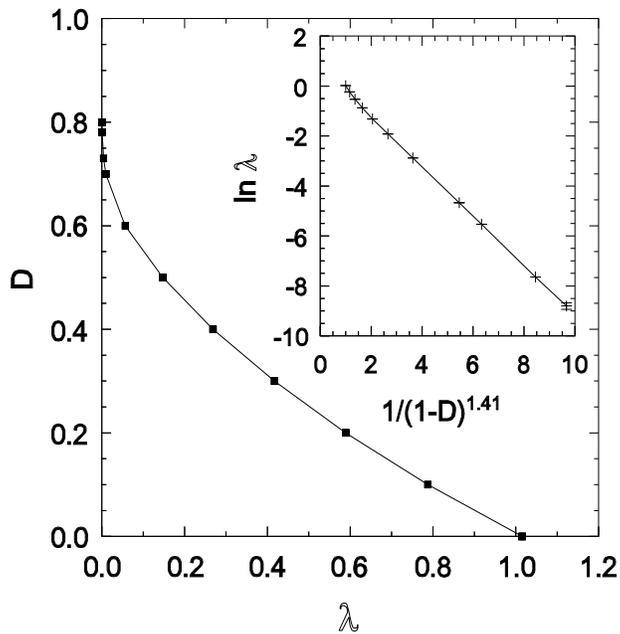}}
\caption {\sf Critical line of the two-dimensional PAM.  Inset:
the same data plotted as $\ln \lambda_c$ versus\\ $1/(1-D)^{1.41}$.}
\label{lc2da}
\end{figure}

\noindent Applied to the data for $D \geq 0.4$, a least-squares procedure yields $\gamma = 1.41(2)$,
$C = 0.984(2)$, and $A = 2.02(2)$.  The good quality of the fit is evident in the inset
of Fig. \ref{lc2da}.  Thus, while a plot of the data on linear scale might suggest
that $\lambda_c \to 0$ at some diffusion rate between 0.8 and 1 (see Fig. \ref{lc2da}, main
graph), our results are in fact consistent with $\lambda_c$ nonzero, though very small,
for diffusion rates between 0.7 and unity.

\subsection{Three dimensions}

\begin{figure}[!h]
\epsfysize=10cm
\epsfxsize=12cm
\centerline{
\epsfbox{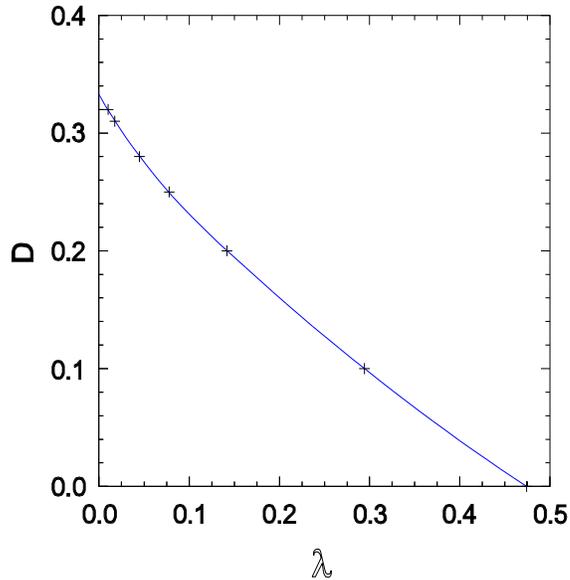}}
\caption {\sf Critical line of the PAM in three dimensions; error bars are smaller than
symbols.  The solid line is a cubic fit to the
data, yielding $D^* = 0.333(3)$.}
\label{lc3d}
\end{figure}

We employed quasistationary simulations to determine $\lambda_c (D)$ for the PAM on the simple cubic lattice.
For relatively small diffusion rates good results are obtained using lattice sizes $L=8$, 16, 24, 36,
and 54.  For diffusion rates greater than about 0.25, however, there are substantial finite-size
effects, and to observe clear signs of DP-like scaling we need to study larger systems ($L=80$ and 120
in addition to the sizes mentioned above).
The results (see Table~\ref{tb:pamss3d} and Fig.~\ref{lc3d}), show that in this case
$\lambda_c$ does fall to zero at a diffusion rate considerably less than unity; extrapolation of
the data to $\lambda = 0$ yields $D^* = 0.333(3)$.  The critical exponents determined via finite-size scaling
analysis, $\beta/\nu_\perp = 1.40(1)$ and $z = 1.94(2)$, are once again in good agreement with literature
values of 1.39(3) and 1.919(4), respectively.
Our study yields the moment ratio value $m = 1.47(1)$ for the three-dimensional models
in the DP universality class; to our knowledge this quantity has not been determined previously.
For $D > D^*$, the particle density is expected to tend to zero linearly with $\lambda$, as the
reproduction rate approaches zero.  We have verified this behavior (down to $\lambda = 10^{-4}$) for
$D=0.8$.

\begin{table}[h]
\begin{center}
\begin{tabular}{|c|c|c|c|c|} \hline
$D$ & $\lambda_c$  \\ \hline\hline
0.0  & 0.47390(5)    \\
0.1  & 0.2943(1)    \\
0.2  & 0.1420(1)    \\
0.25 & 0.07790(5)    \\
0.28 & 0.04487(3)    \\
0.31 & 0.01762(2)    \\
0.32 & 0.0103(1)     \\ \hline
\end{tabular}
\end{center}
\caption{\sf Critical parameters obtained through quasistationary simulations
in three dimensions.}
\label{tb:pamss3d}
\end{table}

\section{Discussion}

We study the phase boundary of the pair annihilation model in the
reproduction rate - diffusion probability ($\lambda$ - $D$) plane.  Our simulation results
are consistent with the theorem proven some time ago by Katori and Konno \cite{katori},
namely that in one and two dimensions, $\lambda_c > 0$ for any $D<1$.
The pair approximation is in conflict with this result, as it predicts that in any number
of dimensions, there is
a diffusion probability $D^* < 1$, above which $\lambda_c = 0$.
In one dimension
the behavior (in simulations) is straightforward, as $\lambda_c \propto 1-D$ for $D \simeq 1$.  In two
dimensions however it is quite subtle, as $\lambda_c$ becomes exponentially small as
$D \to 1$, and a cursory analysis could well give the impression that $\lambda_c$ is
actually zero at some value of $D$ between 0.8 and unity.
Finally in three dimensions
the pair approximation prediction is verified qualitatively; we find $D^* = 0.333(3)$
in this case, while the PA yields $D^* = 1/6$.
Intuitively, the unusual behavior of $\lambda_c(D)$ in two dimensions can be understood
as a consequence of $d=2$ marking a critical dimension for the recurrence of a random walk.
Our simulation results for critical exponents and the moment ratio $m$ are consistent
with the directed percolation values, as expected.
Given the qualitative failure of the pair approximation in one and two dimensions,
it is natural to ask whether approximations using larger clusters would predict
the phase diagram correctly.  This strikes us as unlikely, since cluster
approximations have been found to be insensitive to subtle effects involving diffusion and/or
multiparticle rules in other cases \cite{ben,fontanari,trpcr2009}.

\vspace{1em}

\noindent {\bf Acknowlegdments}
\vspace{1em}
This work was supported by CNPq, Brazil.

\bibliographystyle{apsrev}

\end{document}